\documentclass[prd,twocolumn,superscriptaddress,floatfix,nofootinbib]{revtex4-2}
\pdfoutput=1 % if your are submitting a pdflatex (i.e. if you have
\usepackage[T1]{fontenc} % if needed
\usepackage{amsmath,amssymb,amsfonts,amsthm}
\usepackage{graphicx}
\usepackage[usenames,dvipsnames]{color}
\usepackage{enumitem}
\usepackage{verbatim}
\usepackage[percent]{overpic}
\usepackage{rotating}
\usepackage{hyperref}
\usepackage{array}
\usepackage{mathtools}
\usepackage{lmodern}
\usepackage[normalem]{ulem}
\usepackage{braket}
\usepackage{tensor}
\usepackage{dsfont}
\usepackage{bm}
\usepackage{tikz}

\usepackage{mathrsfs}

\usetikzlibrary{decorations.pathmorphing}
\usepackage{CJKutf8}

\usetikzlibrary{positioning}
\usetikzlibrary{calc,through,backgrounds}

\theoremstyle{definition}
\newtheorem{definition}{Definition}[section]
\newtheorem{theorem}{Theorem}[section]

\newcommand{\kako}[1]{\left( #1 \right)}

\newcommand{\ts}[1]{ _{\text{#1}} }

\newcommand{\erfc}{\text{erfc}}

\allowdisplaybreaks

\DeclareMathOperator{\Tr}{Tr}

\newcommand{\dd}{\text{d}}
\newcommand{\bk}{{\bm{k}}}
\newcommand{\bx}{{\bm{x}}}

\newcommand{\id}{\mathds{1}}
\newcommand{\sx}{\mathsf{x}}
\newcommand{\sy}{\mathsf{y}}

\newcommand{\ii}{\mathsf{i}}

\newcommand{\kk}{|\bm{k}|}
\newcommand{\AAA}{\text{A}}
\newcommand{\BB}{\text{B}}

\begin{document}

\title{Mutual information harvested by uniformly accelerated particle detectors}

%Entangled Detectors Nonperturbatively Harvest Mutual Information
%Entangled Detectors that Gain and Lose Correlations by Interacting a Quantum Field

\author{Manar Naeem}
\email{manar.naeem@uwaterloo.ca}
\affiliation{Department of Physics and Astronomy, University of Waterloo, Waterloo, Ontario, N2L 3G1, Canada}
\affiliation{Institute for Quantum Computing, University of Waterloo, Waterloo, Ontario, N2L 3G1, Canada}

\author{Kensuke Gallock-Yoshimura}
\email{kgallock@uwaterloo.ca} 

\affiliation{Department of Physics and Astronomy, University of Waterloo, Waterloo, Ontario, N2L 3G1, Canada}

%\author{Erickson Tjoa}
%\email{e2tjoa@uwaterloo.ca}
%\affiliation{Department of Physics and Astronomy, University of Waterloo, Waterloo, Ontario, N2L 3G1, Canada}
%\affiliation{Institute for Quantum Computing, University of Waterloo, Waterloo, Ontario, N2L 3G1, Canada}

\author{Robert B. Mann}
\email{rbmann@uwaterloo.ca}
\affiliation{Department of Physics and Astronomy, University of Waterloo, Waterloo, Ontario, N2L 3G1, Canada}
%\affiliation{Perimeter Institute for Theoretical Physics,  Waterloo, Ontario, N2L 2Y5, Canada}

\begin{abstract}
We investigate the mutual information harvesting protocol for two uniformly accelerated particle detectors. 
We numerically show that, while a single detector responds as if it is immersed in a thermal bath, the quantum mutual information between two accelerating detectors behaves differently than that of two inertial detectors in a thermal bath. 
This is  due to the fact that while the   Wightman function along the trajectory of a single uniformly accelerating detector is the same as that of  as a detector in a thermal bath, a pair of detectors in the same respective cases will have different Wightman functions. 
\end{abstract}

\maketitle
\flushbottom

\section{Introduction}
Quantum information theory has become one of the most attractive fields of study in recent decades, with many protocols such as quantum teleportation,  quantum key distribution and other cryptographic protocols being developed for practical purposes.  
Although these protocols are quite remarkable, they are generally formulated in non-relativistic settings. 
However in fully realistic scenarios, the physics of quantum information processing must obey relativistic constraints. 
Relativistic quantum information theory takes such  constraints into account, and in general is concerned with the interplay between relativistic effects and quantum information tasks.

It is common in relativistic quantum information to employ a localized qubit model known as Unruh-DeWitt (UDW) particle detector \cite{Unruh1979evaporation, DeWitt1979}, which interacts with quantum fields in spacetime. 
By using such a qubit model, one can examine, for example, the channel capacity of quantum fields \cite{Cliche2010channel, jonsson2017quantum, Landulfo2016magnus1, Simidzija2020transmit, Tjoa.RQC} and the relativistic quantum teleportation protocol \cite{PhysRevLett.91.180404, Landulfo2009suddendeath}. 
% Quantum information theory has become one of the most attractive studies in the past decades. 
% In particular, the quantum teleportation protocol and quantum key distribution utilize quantum entanglement, which is a characteristic correlation in quantum theory. 

In recent years, the \textit{entanglement harvesting protocol} has been extensively explored. 
The protocol is the following: 
suppose two observers with a UDW detector are in a (flat or curved) spacetime on which a quantum field is defined. 
Assuming the detectors are initially uncorrelated, by locally interacting with the quantum field, the detectors become entangled after the interaction \cite{Valentini1991nonlocalcorr, reznik2003entanglement, reznik2005violating}. 
This is true even when the observers are causally disconnected because the quantum field is already in an entangled state \cite{summers1985bell, summers1987bell}.

More generally, the detectors are extracting correlations from the quantum field, and so we can refer to the protocol as the 
\textit{correlation harvesting protocol}. 
These correlations can be entanglement, mutual information, or quantum discord \cite{Henderson2001correlations, Zurek2001discord, Brown2013Amplification, SahuSabotage}. 
The amount of correlation extracted is influenced by the geometry of spacetime \cite{pozas2015harvesting, smith2016topology, kukita2017harvesting, henderson2018harvesting, ng2018AdS, cong2020horizon, FinnShockwave} and the states of motion of the detectors \cite{salton2015acceleration, FooSuperpositionTrajectory, Diki.inertial}.

There have been numerous studies of the effects of temperature on correlation harvesting. 
The earliest investigation was carried out by Ver Steeg and Menicucci \cite{Steeg2009}. 
They found that, while a single detector in de Sitter spacetime responds as if it is in a thermal bath in Minkowski spacetime, the harvested entanglement between two detectors differs in de Sitter spacetime and a thermal bath. 
%They examined harvested entanglement in de Sitter spacetime and showed that, while a single detector responds as if it is in a thermal bath in Minkowski spacetime, the harvested entanglement between two detectors can distinguish  from it. 
Since then, the temperature dependence of correlations in various scenarios has been analyzed. 
A basic scenario consists of two inertial detectors interacting with a field in a thermal state. 
In \cite{Brown2013harmonic, simidzija2018harvesting} it was found that the amount of entanglement decreases with temperature whereas quantum mutual information increases monotonically.

One can also think of two uniformly accelerating detectors in Minkowski spacetime. 
For a single particle detector, it is widely known that a  detector undergoing uniform acceleration $a$
experiences a thermal bath at temperature $T\ts{U}=\hbar a/2\pi k\ts{B}c$. 
Such a phenomenon is known as the Unruh effect \cite{Unruh1979evaporation} and the temperature $T\ts{U}$ is called the Unruh temperature. 
In this case, the field is in the Minkowski vacuum state instead of a thermal state, but due to its motion the detector responds in the same manner as if it were in the thermal quantum field. 
The harvested entanglement by uniformly accelerated detectors, however, shows a different temperature dependence in contrast to that of a thermal state; entanglement is enhanced at relatively smaller temperatures, then decreases and drops to 0 as the detectors experience hotter Unruh temperature \cite{Liu:2021dnl}. 
The effect of acceleration on entanglement has been studied in other papers as well \cite{FuentesAliceFalls, salton2015acceleration, AlsingDiracFields, Alsing2012review, Zhang.harvesting.circular, Liu.acceleration.vs.thermal}.

One could ask if the behavior induced by acceleration is present in black hole spacetimes, since the Rindler metric is similar to the Schwarzschild one near its event horizon. 
Investigations of entanglement harvesting outside of black holes, while indicating there are indeed common features, have
also found some surprising results, including the presence of an entanglement shadow near the horizon
\cite{henderson2018harvesting} and the amplification of harvested entanglement for near-extremal rotating black holes
\cite{robbins2020entanglement}. Other kinds of black hole spacetimes have been shown to exhibit other novel features
\cite{Henderson2019anti-hawking,Robbins-Anti-Hawking,Henderson:2022oyd,robbins2020entanglement,DeSouzaCampos:2020ddx,Tjoa2020vaidya, Ken.Freefall.PhysRevD.104.025001}. 
In general, correlations steeply decline 
whereas excitation responses of detectors tend to increase as one of the static detectors is placed increasingly close to the horizon,
where the black hole temperature is extremely high. 

The aforementioned studies (inertial detectors in a thermal quantum field, uniformly accelerated detectors in Minkowski vacuum, and static detectors in a black hole spacetime) have one thing in common: a single detector experiences a thermal bath at the corresponding temperature. 
That is, the response of the detector against temperature is the same in all these cases.  Yet the entanglement harvested between two detectors exhibits different temperature dependence, distinguishing the two scenarios from each other. 

Quantum mutual information has been considerably less explored. 
The importance of studying quantum mutual information arises from its applications in multiple areas such as quantum communication~\cite{Qcommunication} and quantum error correction~\cite{Ogawa.error.correction}.
A recent study~\cite{Kendra.BTZ} explicitly examined the relationship between quantum mutual information and the Ba\~{n}ados-Teitelboim-Zanelli (BTZ) black hole temperature. 
Unlike the case for entanglement harvesting, harvested mutual information vanished only when a (static) detector was placed arbitrarily close to the event horizon. 
It was the extremity of Hawking and Unruh effects near the horizon that inhibited the detectors from harvesting.
In other words, the mutual information between two static detectors decreases, vanishing at high black hole temperatures.

\begin{figure*}[t]
\centering
\includegraphics[width=\textwidth]{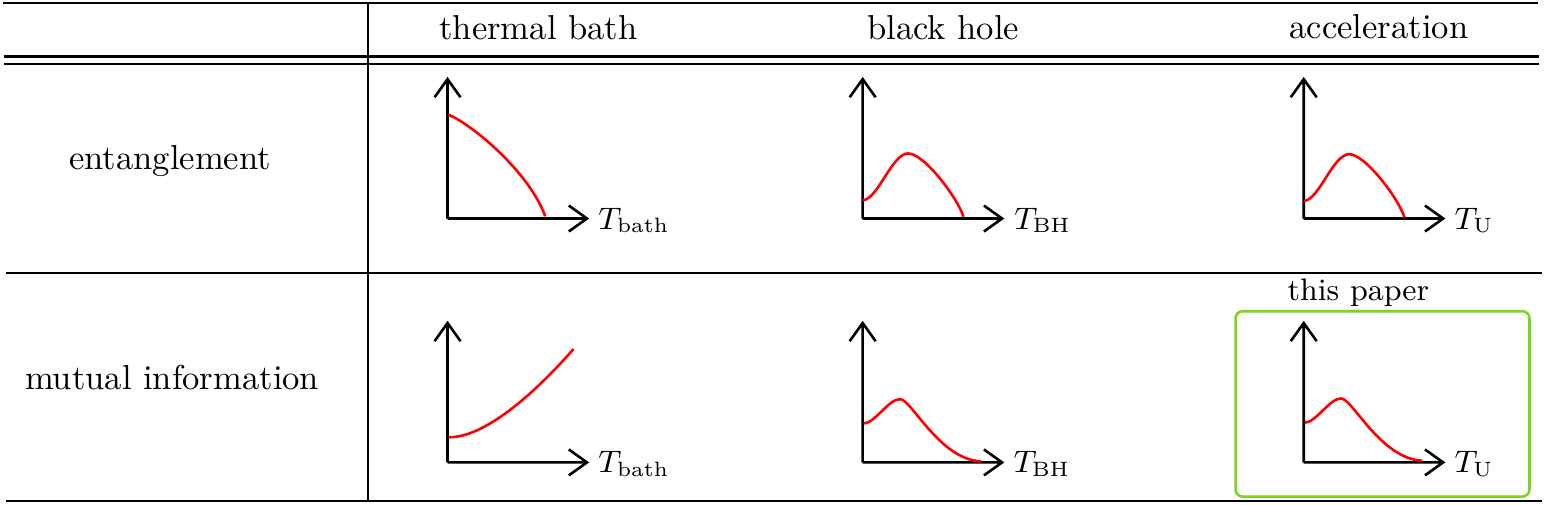}
\caption{A summary of past results on the temperature dependence of correlation harvesting. 
Our results provide the case of mutual information with accelerating detectors as depicted in the lower-right corner.}
\label{fig:past results}
\end{figure*}

This paper aims to complete the picture of the temperature dependence of correlation harvesting. 
We illustrate in Fig.~\ref{fig:past results} the qualitative distinctions between the various scenarios. 
Specifically we analyze the quantum mutual information harvested between two uniformly accelerating detectors. 
We shall consider three configurations of accelerating detector pairs -- parallel, anti-parallel, and perpendicular  \cite{Liu:2021dnl} -- and compare mutual information harvesting and entanglement harvesting with changing acceleration.

We find that harvested quantum mutual information behaves in a broadly qualitatively similar way to harvested entanglement. 
For all three configurations of acceleration, both harvested mutual information and entanglement are
 enhanced at low  temperatures, but become extinguished  in the high Unruh temperature limit.   
We will comment on how to understand  differences in the temperature dependence of correlation harvesting illustrated in Fig.~\ref{fig:past results} from the Wightman function perspective. 
Based on this observation, we infer the temperature dependence of harvested correlations in de Sitter spacetime in \cite{Steeg2009, Nambu.correlations.inflation}.

Our paper is organized as follows. 
By introducing the UDW detector model and three acceleration configurations in Sec.~\ref{sec:setup}, we numerically evaluate quantum mutual information in Sec.~\ref{sec: results}, followed by Conclusion in Sec. \ref{sec: conclusion}. 
Throughout this paper, we use the units $\hbar = k\ts{B}=c=1$ and the signature $(-,+,+,+)$, as well as $\sx\coloneqq x^\mu$, an event in coordinates $x^\mu$.

\section{Setup}
\label{sec:setup}

\subsection{Density matrix}
\label{subsec:density matrix}

A (monopole type) UDW detector is a two-level quantum system that can be considered as a qubit locally coupled to a quantum field. 
We denote the ground and excited states for detector-$j \in \{\AAA, \BB\}$ by $\ket{g_j}$ and $\ket{e_j}$, respectively, with the energy gap $\Omega_j$ between them in the detector's reference frame. 

Let us introduce two UDW particle detectors A and B that   interact with a quantum scalar field $\hat \phi$. 
The dynamics of the detectors is governed by a Hamiltonian. 
Assuming the size of the detectors is negligible and each detector has its own proper time $\tau_j$, the interaction Hamiltonian in the interaction picture is given by
\begin{align}
    \hat H_j^{ \tau_j } ( \tau_j )
    &=
        \lambda_j \chi_j(\tau_j) \hat \mu_j(\tau_j) 
        \otimes \hat \phi(\sx_j(\tau_j))\,.~j\in \{ \AAA, \BB \}
\end{align}
The quantity $\lambda_j \chi_j(\tau_j)$ specifies the time-dependence of coupling between detector-$j$ and the field, and so $\chi_j(\tau_j)$ is called a switching function. 
The operator $\hat \mu_j(\tau_j) $ is the so-called monopole moment, which describes the internal dynamics of a detector and is given by 
\begin{align}
    \hat \mu_j(\tau_j) 
    &=
        \ket{e_j} \bra{g_j} e^{ \ii \Omega_j \tau_j }
        +
        \ket{g_j} \bra{e_j} e^{ -\ii \Omega_j \tau_j }\,.
\end{align}
$\hat \phi(\sx_j(\tau_j))$ is the field operator along detector-$j$'s trajectory. 
In this sense, the detector locally couples to the field at the point where the detector is located. 
The superscript on the Hamiltonian $\hat H_j^{ \tau_j } ( \tau_j )$ indicates that it is the generator of time-translation with respect to the proper time $\tau_j$.

The total interaction Hamiltonian, $\hat H\ts{I}^t(t)$, that describes both detectors A and B is then given by
\begin{align}
    \hat H\ts{I}^t(t)
    &=
        \dfrac{\dd \tau\ts{A}}{\dd t} 
        \hat H\ts{A}^{ \tau\ts{A} }\big( \tau\ts{A}(t) \big)
        +
        \dfrac{\dd \tau\ts{B}}{\dd t} 
        \hat H\ts{B}^{ \tau\ts{B} }\big( \tau\ts{B}(t) \big) \,, 
\end{align}
where the Hamiltonian $\hat H\ts{I}^t(t)$ is now a generator of time-translation with respect to the common time $t$ (e.g., Minkowski time). 
The time-evolution operator $\hat U\ts{I}$ can be written by using a time-ordering symbol $\mathcal{T}_t$ with respect to $t$ \cite{EMM.Relativistic.quantum.optics,Tales2020GRQO}: 
\begin{align}
    \hat U\ts{I}
    &=
        \mathcal{T}_t 
        \exp 
        \kako{
            -\ii \int_{\mathbb{R}} \dd t\,\hat H\ts{I}^t(t)
        } .
\end{align}
Assuming that the coupling strength $\lambda$ is small, the Dyson series expansion of $\hat U\ts{I}$ reads: 
\begin{subequations}
\begin{align}
    \hat U\ts{I}
    &=
        \id + \hat U\ts{I}^{(1)} + \hat U\ts{I}^{(2)} + \mathcal{O}(\lambda^3)\,,\\
    \hat U\ts{I}^{(1)}
    &=
        -\ii \int_{-\infty}^\infty \dd t\,\hat H\ts{I}^t(t)\,,\\
    \hat U\ts{I}^{(2)}
    &=
        - \int_{-\infty}^\infty \dd t_1
        \int_{-\infty}^{t_1} \dd t_2\,
        \hat H\ts{I}^t(t_1) \hat H\ts{I}^t(t_2)\,.
\end{align}
\end{subequations}

Let us now obtain the density matrix for the detectors up to the second order in $\lambda$. 
We assume that the initial states of the detectors and the field are in the ground $\ket{g\ts{A}} \otimes \ket{g\ts{B}}$ and vacuum $\ket{0}$ states respectively, and uncorrelated.  The rationale for this assumption is twofold.  First, it is a natural assumption to make in the lab, as it is straightforward to initialize systems to be in their ground states. Second, this assumption allows us to study harvesting of vacuum correlations  without contaminants from other sources.  It is certainly possible to relax the assumption of an initially uncorrelated state, and recently studies of entanglement harvesting in this context have been carried out
\cite{Chowdhury:2021ieg,Bhattacharya:2022ahn}.

We therefore take
the initial state $\rho_0$ of the total system to be
\begin{align}
    \rho_0
    &=
        \ket{g\ts{A}} \bra{g\ts{A}}
        \otimes 
        \ket{g\ts{B}} \bra{g\ts{B}}
        \otimes 
        \ket{0}\bra{0}\,.
\end{align}
The final total density matrix $\rho\ts{tot}$ after the interaction reads
\begin{align}
    \rho\ts{tot}
    &=
        \hat U\ts{I} \rho_0 \hat U\ts{I}^\dag \notag \\
    &=
        \rho_0 
        + 
        \rho^{(1,1)}
        +
        \rho^{(2,0)}
        +
        \rho^{(0,2)}
        +
        \mathcal{O}(\lambda^4)\,,
\end{align}
where $\rho^{(i,j)}=\hat U^{(i)} \rho_0 \hat U^{(j)\dagger}$. 
Note that all the odd-power terms of $\lambda$ vanish \cite{pozas2015harvesting}, in the final density matrix of the detectors upon tracing out the field degree of freedom: $\rho\ts{AB}=\Tr_\phi[\rho\ts{tot}]$. 
In the basis $\{ \ket{g\ts{A} g\ts{B}} , \ket{g\ts{A} e\ts{B}}, \ket{e\ts{A} g\ts{B}}, \ket{e\ts{A} e\ts{B}} \}$, $\rho\ts{AB}$ is known to be
\begin{align}
    \rho\ts{AB}
    &=
        \left[
        \begin{array}{cccc}
        1-\mathcal{L}\ts{AA}-\mathcal{L}\ts{BB} &0 &0 &\mathcal{M}^*  \\
        0 &\mathcal{L}\ts{BB} &\mathcal{L}\ts{AB}^* &0  \\
        0 &\mathcal{L}\ts{AB} &\mathcal{L}\ts{AA} &0  \\
        \mathcal{M} &0 &0 &0 
        \end{array}
        \right]
        + \mathcal{O}(\lambda^4)
        \label{eq:density matrix}
\end{align}
where
\begin{align}
    \mathcal{L}_{ij}
    &=
        \lambda^2
        \int_{\mathbb{R}} \dd \tau_i
        \int_{\mathbb{R}} \dd \tau_j'\,
        \chi_i(\tau_i) \chi_j(\tau_j')
        e^{ -\ii \Omega (\tau_i - \tau_j') } \notag \\
        &\qquad\qquad\qquad \qquad\times 
        W\big( \sx_i(\tau_i), \sx_j(\tau_j') \big)\,, \\
    \mathcal{M}
    &=
        -\lambda^2
        \int_{\mathbb{R}} \dd \tau\ts{A}
        \int_{\mathbb{R}} \dd \tau\ts{B}\,
        \chi\ts{A}(\tau\ts{A}) \chi\ts{B}(\tau\ts{B})
        e^{ \ii \Omega (\tau\ts{A} + \tau\ts{B}) } \notag \\
        &\hspace{5mm}\times 
        \big[ 
            \Theta \big( t(\tau\ts{A}) - t(\tau\ts{B}) \big)
            W \big( \sx\ts{A}(\tau\ts{A}), \sx\ts{B}(\tau\ts{B}) \big) \notag \\
            &\hspace{1cm}
            +
            \Theta \big( t(\tau\ts{B}) - t(\tau\ts{A}) \big)
            W \big( \sx\ts{B}(\tau\ts{B}), \sx\ts{A}(\tau\ts{A}) \big)
        \big]\,,
\end{align}
where $\Theta(t)$ is the Heaviside step function and the quantity $W(\sx, \sy)\coloneqq \bra{0} \hat\phi(\sx) \hat \phi(\sy) \ket{0}$ is the Wightman function. 
The elements $\mathcal{L}_{jj},\,j\in \{\AAA, \BB\}$ are known as transition probabilities (or responses) from the ground $\ket{g_j}$ to excited $\ket{e_j}$ states. 
The off-diagonal elements $\mathcal{M}$ and $\mathcal{L}\ts{AB}$ contribute to entanglement and mutual information, respectively. 
%It is worth knowing that the elements satisfy $\mathcal{L}\ts{AA}\mathcal{L}\ts{BB}\geq |\mathcal{L}\ts{AB}|^2$ \cite{smith2016topology}. 

Mutual information $I\ts{AB}$ between detectors A and B up to   second order in $\lambda$ is \cite{pozas2015harvesting}
\begin{align}
    I\ts{AB}
    &=
        \mathcal{L}_+ \ln \mathcal{L}_+
        + 
        \mathcal{L}_- \ln \mathcal{L}_- \notag \\
        &\hspace{5mm}-
        \mathcal{L}\ts{AA} \ln \mathcal{L}\ts{AA}
        -
        \mathcal{L}\ts{BB} \ln \mathcal{L}\ts{BB}
        + \mathcal{O}(\lambda^4)
        \,,
\end{align}
where 
\begin{align}
    \mathcal{L}_\pm
    &\coloneqq
        \dfrac{1}{2}
        \kako{
            \mathcal{L}\ts{AA}
            +
            \mathcal{L}\ts{BB}
            \pm 
            \sqrt{ (\mathcal{L}\ts{AA}-\mathcal{L}\ts{BB})^2 + 4 |\mathcal{L}\ts{AB}|^2 }
        }. \label{eq:Lpm}
\end{align}
Mutual information vanishes when $|\mathcal{L}\ts{AB}|=0$. 
In particular, from the condition $\mathcal{L}\ts{AA}\mathcal{L}\ts{BB}\geq |\mathcal{L}\ts{AB}|^2$ \cite{smith2016topology}, if one of the transition probabilities satisfies $\mathcal{L}_{jj}=0$ then $|\mathcal{L}\ts{AB}|=0$, thereby $I\ts{AB}=0$. 

Throughout this paper, we use a Gaussian switching function
\begin{align}\label{Gswitch}
    \chi_j(\tau_j)
    &=
        e^{ -\tau_j^2/2\sigma^2 }\,,
\end{align}
where $\sigma$ is a typical interaction duration. 
We will write and evaluate all quantities in units of $\sigma$. 
For example, acceleration constant $a$, energy gap $\Omega$, and the proper time $\tau$ respectively  become $a\sigma, \Omega \sigma$, and $\tau/\sigma$.

\begin{figure*}[t]
\centering
\includegraphics[width=\textwidth]{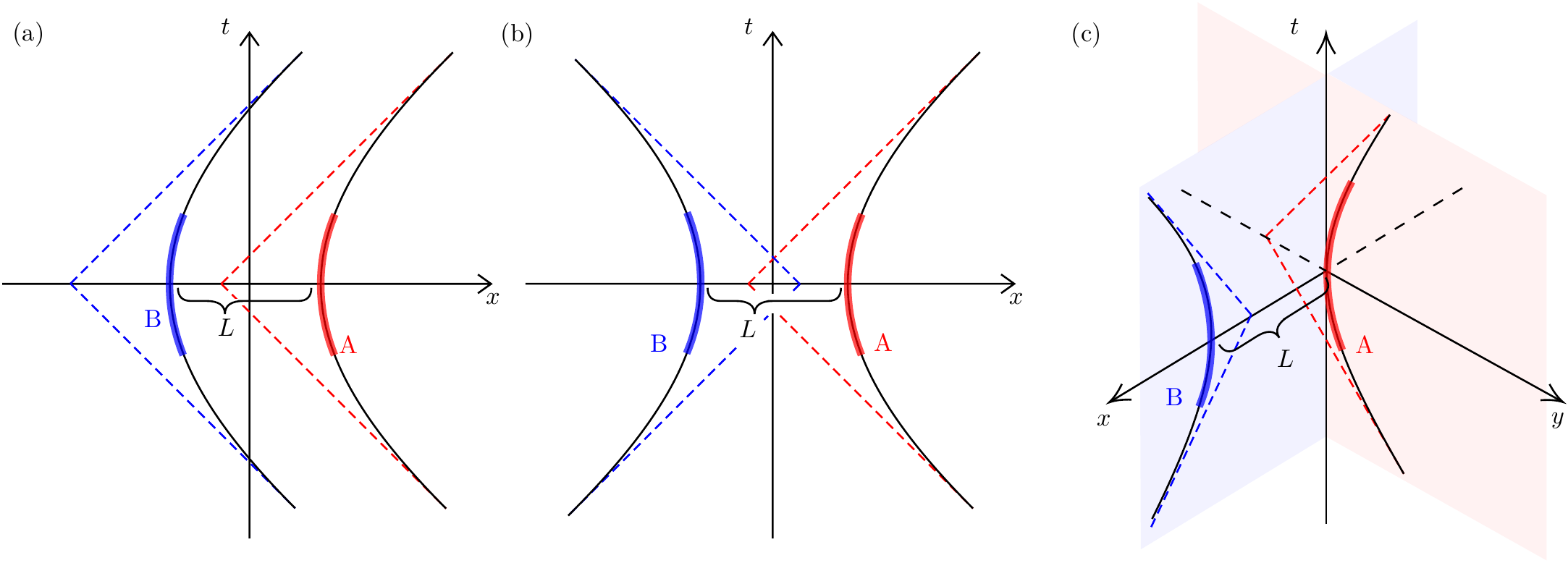}
\caption{Three configurations of acceleration: (a) parallel, (b) anti-parallel, and (c) perpendicular. 
The red and blue stripes indicate the interaction duration of detectors A and B, respectively. 
In all cases, the detectors are separated by $L$ at $t=0$, at which their Gaussian switching peaks. 
Note that the separation in the parallel configuration is $L$ for all times.}
\label{fig:acceleration scenarios}
\end{figure*}

\subsection{Detectors' trajectories}
Let us now restrict ourselves to uniformly accelerating detectors in $(3+1)$-dimensional Minkowski spacetime. 
In particular, we employ a massless quantum scalar field $\hat \phi(\sx)$ that obeys the Klein-Gordon equation and we assume that the field is minimally coupled. 
One can perform a mode expansion as 
\begin{align}
    \hat \phi(\sx)
    &=
        \int \dfrac{\dd^3 k}{ \sqrt{ (2\pi)^3 2\kk } }
        \kako{
            \hat a_{\bk} e^{ -\ii \kk t + \ii \bk \cdot \bx }
            + \text{h.c.}
        }\,,
\end{align}
with the Minkowski vacuum $\ket{0}$ satisfying $\hat a_\bk \ket{0}=0$ for all $\bk$. 
The creation and annihilation operators obey the canonical commutation relations, 
\begin{subequations}
\begin{align}
    &[\hat a_\bk , \hat a_{\bk'}^\dag]
    =
        \delta^{(3)}(\bk - \bk')\,,\\
    &[\hat a_\bk , \hat a_{\bk'}]=0\,,
    \quad 
    [\hat a_\bk^\dag , \hat a_{\bk'}^\dag]=0\,.
\end{align}
\end{subequations}
Then the Wightman function in the Minkowski vacuum state $\ket{0}$ is known to be 
\begin{align}
    W(\sx, \sx')
    &=
        -\dfrac{1}{4\pi^2}
        \dfrac{1}{(t-t'-\ii \epsilon)^2 - |\bx - \bx'|^2}\,,\label{eq:Wightman in Minkowski}
\end{align}
where $\epsilon$ is a UV regulator.

To evaluate the elements in the density matrix $\rho\ts{AB}$, one needs to specify the trajectories of detectors A and B in this Wightman function. 
In what follows, we will consider three different acceleration scenarios: parallel, anti-parallel, and perpendicular.

\subsubsection{Parallel acceleration}

In this scenario   a pair of UDW detectors A and B
accelerating in the same direction along $x$, separated by distance $L$ for all times, as shown in Fig.~\ref{fig:acceleration scenarios}(a). 
 The detectors' trajectories can be written as 
\begin{align}
    \sx\ts{A}
    &=
        \Big\{ 
            t=\dfrac{1}{a} \sinh (a\tau\ts{A}),
            x= \dfrac{1}{a} [\cosh (a\tau\ts{A})-1]+\dfrac{L}{2}, \notag \\
            &\hspace{1cm}y=z=0 
        \Big\}\,, \\
    \sx\ts{B}
    &=
    \Big\{
    t=\dfrac{1}{a} \sinh (a\tau\ts{B}),
    x=\dfrac{1}{a} [\cosh (a\tau\ts{B})-1]- \dfrac{L}{2}, \notag \\
    &\hspace{1cm}y=z=0 
    \Big\}\,. 
\end{align}
By substituting these trajectories into the Wightman function \eqref{eq:Wightman in Minkowski}, one can evaluate the elements in the density matrix \eqref{eq:density matrix} along the trajectories. 
The Wightman function $W_a(\sx, \sy)$ becomes 
\begin{widetext}
    \begin{align}
    W_a(\sx_j(\tau_j), \sx'_j(\tau'_j))
    &= 
        -\dfrac{a^2}{16\pi^2}
        \dfrac{ 1 }{ \sinh^2[ a(\tau_j - \tau_j')/2 - \ii \epsilon ] } \,,
        \hspace{0.5cm} j \in \{ \AAA, \BB \} \label{eq:para single Wightman} \\
    W_a(\sx\ts{A}(\tau\ts{A}), \sx\ts{B}(\tau\ts{B}))
    &=
        -\dfrac{a^2}{4\pi^2}
        \dfrac{ 1 }{ [ \sinh(a\tau\ts{A}) - \sinh(a\tau\ts{B}) -\ii \epsilon ]^2 - |\cosh (a\tau\ts{A}) - \cosh (a\tau\ts{B}) + a L|^2 }\,.\label{eq:para two detector Wightman}
\end{align}
It is worth pointing out that, for the Gaussian switching \eqref{Gswitch}, the transition probability $\mathcal{L}_{jj}$ can be simplified to \cite{Liu:2021dnl} 
\begin{align}
    \mathcal{L}_{jj}
    &=
        \dfrac{\lambda^2}{4\pi}
        [
            e^{ -\Omega^2 \sigma^2 }
            -\sqrt{\pi} \Omega \sigma \erfc(\Omega \sigma)
        ]  +\dfrac{\lambda^2 a \sigma}{ 4 \pi^{3/2} }
        \int_0^\infty \dd s \,
        \dfrac{ \cos (\beta s) e^{ -\alpha s^2 } (\sinh^2 s - s^2) }{ s^2 \sinh^2 s } \,,\label{eq:transition prob}
\end{align}
\end{widetext}
where $\beta \equiv 2 \Omega /a$ and $\alpha \equiv 1/(a\sigma)^2$. 
Note that the first term in $\mathcal{L}_{jj}$ is the transition probability of a detector at rest in $(3+1)$-dimensional Minkowski spacetime.

\begin{figure*}[t]
\centering
\includegraphics[width=\textwidth]{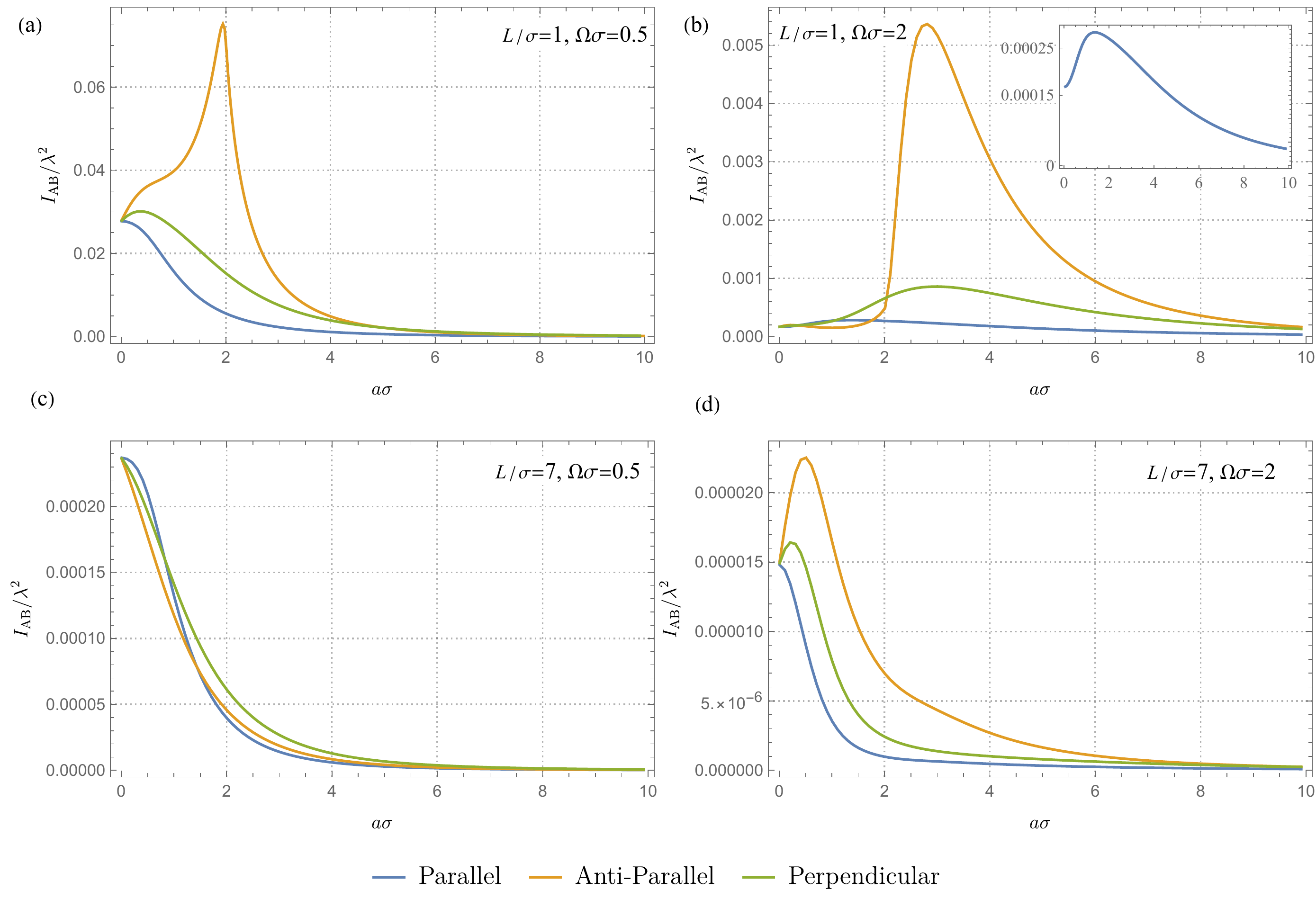}
\caption{Mutual Information as a function of acceleration $a \sigma$ in three acceleration scenarios (parallel, anti-parallel, and perpendicular). 
(a) $L/\sigma=1, \Omega \sigma = 0.5$, (b) $L/\sigma=1, \Omega \sigma=2$, (c) $L/\sigma=7, \Omega \sigma=0.5$, and (d) $L/\sigma=7, \Omega \sigma=2$. 
}
\label{fig:MI different accelerations}
\end{figure*}

\subsubsection{Anti-parallel acceleration}

The anti-parallel configuration, shown in Fig.~\ref{fig:acceleration scenarios}(b), is the case where two detectors accelerate toward each other and after momentarily stopping (at which point the detector separation is $L$), they accelerate away. 
Unlike the parallel acceleration configuration, the distance between the detectors is not fixed. 

The trajectories are given by
\begin{align}
    \sx\ts{A}
    &=
        \Big\{ 
            t=\dfrac{1}{a} \sinh (a\tau\ts{A}),
            x= \dfrac{1}{a} [\cosh (a\tau\ts{A})-1]+\dfrac{L}{2}, \notag \\
            &\hspace{1cm}y=z=0 
        \Big\}\,, \\
    \sx\ts{B}
    &=
    \Big\{
    t=\dfrac{1}{a} \sinh (a\tau\ts{B}),
    x=\dfrac{-1}{a} [\cosh (a\tau\ts{B})-1]- \dfrac{L}{2}, \notag \\
    &\hspace{1cm}y=z=0 
    \Big\}\,.
\end{align}
While the Wightman function along both the trajectories, $W(\sx\ts{A}, \sx\ts{B})$, differs from \eqref{eq:para two detector Wightman} and therefore $\mathcal{L}\ts{AB}$ and $\mathcal{M}$ in the density matrix \eqref{eq:density matrix}, the transition probability $\mathcal{L}_{jj}$ is the same as \eqref{eq:transition prob}. 
Note that the detectors, as long as $L$ is small, can in general communicate with each other by exchanging field quanta when the detectors are lightlike separated.

\subsubsection{Perpendicular acceleration}

As depicted in Fig.~\ref{fig:acceleration scenarios}(c), the perpendicular acceleration configuration is similar to the anti-parallel configuration, but now detectors are traveling along different axes $x$ and $y$. 
That is, the two detectors accelerate toward and away from each other with a $90^\circ$ angle. 
The minimum distance between them at which they stop momentarily is $L$. 
The trajectories are 
\begin{align}
    \sx\ts{A}
    &=
        \Big\{ 
            t=\dfrac{1}{a} \sinh (a\tau\ts{A}),
            y= \dfrac{1}{a} [\cosh (a\tau\ts{A})-1], \notag \\
            &\hspace{1cm}x=z=0 
        \Big\}\,, \\
    \sx\ts{B}
    &=
    \Big\{
    t=\dfrac{1}{a} \sinh (a\tau\ts{B}),
    x=\dfrac{1}{a} [\cosh (a\tau\ts{B})-1]+L, \notag \\
    &\hspace{1cm}y=z=0 
    \Big\}\,.
\end{align}
Note again that the transition probability $\mathcal{L}_{jj}$ of each detector is the same as \eqref{eq:transition prob}.

\begin{figure*}[t]
\centering
\includegraphics[width=\textwidth]{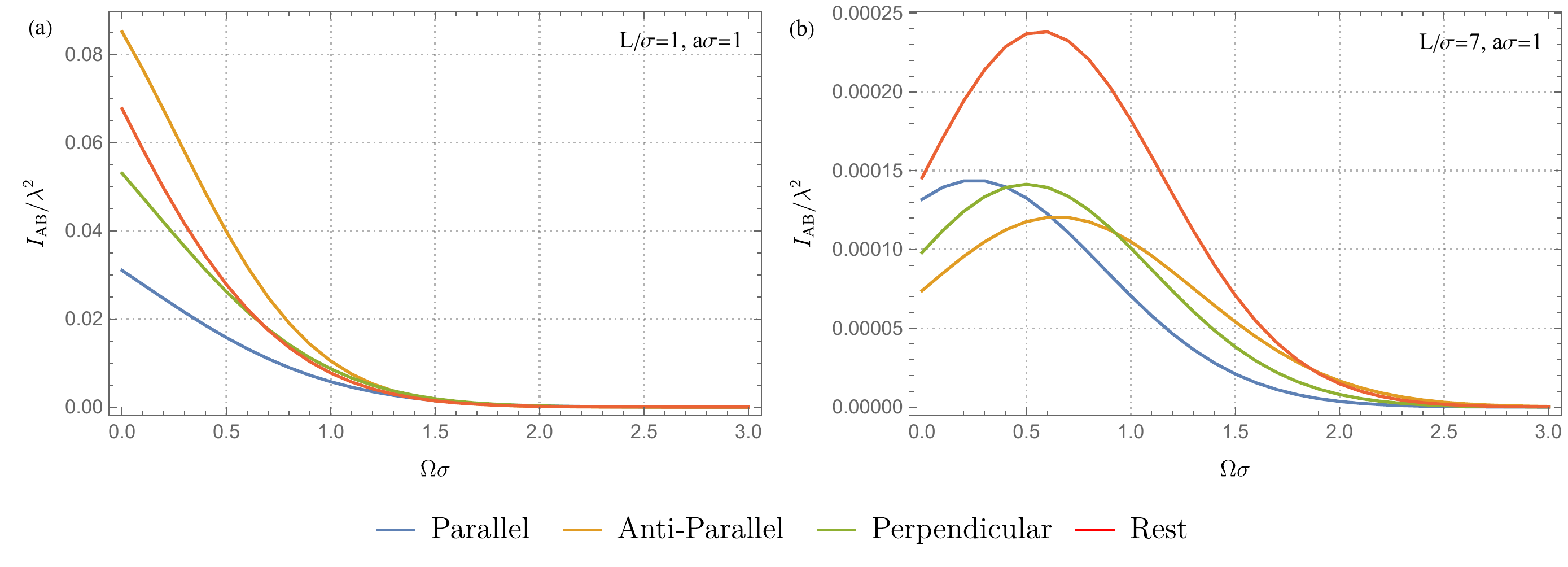}
\caption{Mutual Information as a function of energy gap $\Omega \sigma$ in three acceleration scenarios (parallel, anti-parallel, and perpendicular) with $a \sigma = 1$  and  (a) $L/\sigma=1$ , (b) $L/\sigma=7$. 
The red curve represents the harvested mutual information by two inertial detectors in the Minkowski vacuum, which corresponds to $a\sigma=0$. 
}
\label{fig:MI different Omega}
\end{figure*}

\section{Results}\label{sec: results}

\subsection{Temperature dependence}

Figure~\ref{fig:MI different accelerations} shows the effect of acceleration on mutual information harvesting for each of the scenarios in
Fig.~\ref{fig:acceleration scenarios}, plotting mutual information $I\ts{AB}$ as a function of acceleration $a\sigma$ (which is proportional to the Unruh temperature $T\ts{U}=a/2\pi$). The diagrams depict  different energy gaps $\Omega$ and   detector separations $L$ at $t=0$. 
Figures~\ref{fig:MI different accelerations}(a) and (b) depict $I\ts{AB}$ with $\Omega \sigma =0.5$ and 2, respectively, when the separation is small ($L/\sigma=1$),  whereas (c) and (d) have a large separation: $L/\sigma =7$ for the same two gaps. 
Note that the effect of communication between the two detectors is negligible when $L/\sigma=7$, which suggests that the harvested mutual information predominantly comes from preexisting entanglement in the vacuum state of the field. 

We see that high acceleration suppresses mutual information harvesting in all three acceleration scenarios regardless of the energy gap $\Omega\sigma$ and  separation $L/\sigma$. 
This characteristic property of mutual information can be explained as follows. 
Since the two detectors have the same transition probabilities, $\mathcal{L}\ts{AA}=\mathcal{L}\ts{BB}\equiv P$, then
  $\mathcal{L}_{\pm}$ in \eqref{eq:Lpm} becomes 
\begin{align}
    \mathcal{L}_\pm
    &=
        P \pm |\mathcal{L}\ts{AB}|\,.
\end{align}
The reason that  $I\ts{AB}$ vanishes at high acceleration (or equivalently,  high temperatures $T\ts{U} \to \infty$) is that the transition probability $P$ monotonically increases with $a\sigma$ while $|\mathcal{L}\ts{AB}|$ remains small, which leads to $P \gg |\mathcal{L}\ts{AB}|$ and so $\mathcal{L}_\pm \approx P$. 
Thus, the mutual information $I\ts{AB} \approx 0$. 
However for a thermal bath (Fig.~\ref{fig:past results})  \cite{simidzija2018harvesting},   the mutual information between two inertial detectors increases with $T$ because both $P$ and $|\mathcal{L}\ts{AB}|$ increase with temperature $T$; consequently the mutual information monotonically increases with $T$.

By contrast, small acceleration seems to affect mutual information harvesting differently depending on the type of acceleration, energy gap, and detectors' separation. 
In the case of small energy gap and small detector separation, shown in Fig.~\ref{fig:MI different accelerations}(a), we find that small acceleration enhances mutual information for anti-parallel and perpendicular configurations with higher harvested mutual information in the anti-parallel scenario, while harvested mutual information monotonically decreases with $a\sigma$ in the parallel acceleration case.
Nevertheless, as the energy gap $\Omega \sigma$ changes from 0.5 to 2, the acceleration dependence of $I\ts{AB}$ changes, as shown in Fig.~\ref{fig:MI different accelerations}(b). 
In particular, the parallel acceleration case no longer monotonically decreases with $a\sigma$, and smaller acceleration could enhance mutual information harvesting. 
This is also true for $L/\sigma=7$ in Figs.~\ref{fig:MI different accelerations}(c) and (d).
We also examine how harvested mutual information changes with the energy gap by plotting $I\ts{AB}$ as a function of $\Omega\sigma$ in Fig.~\ref{fig:MI different Omega}. 
Here, we fix the value of acceleration to be $a\sigma=1$, and plot the energy gap dependence when $L/\sigma=1$ and 7 in Figs.~\ref{fig:MI different Omega}(a) and (b), respectively. 
For entanglement harvesting reported in \cite{Liu:2021dnl}, any accelerating detectors (as well as inertial detectors) with small energy gaps cannot extract entanglement when the detector separation $L$ is large. 
However, this is not the case for mutual information; we find that for both $L/\sigma=1$ and 7 in Fig.~\ref{fig:MI different Omega}, mutual information $I\ts{AB}$ is nonvanishing near $\Omega=0$, which suggests that the harvested correlation with small $\Omega$ is either classical correlation or nondistillable entanglement.

\subsection{Comparison to previous studies}

As we have shown in the previous section, harvested quantum mutual information $I\ts{AB}$ behaves in a manner similar to harvested entanglement  \cite{Liu:2021dnl}. 
From Fig.~\ref{fig:past results}, we can now discuss how different the temperature dependence among various scenarios is. 
Here, we focus on the difference between the accelerating detector scenarios and the thermal bath scenario in \cite{Brown2013harmonic, simidzija2018harvesting}.

Let us first review the thermal state of the scalar field. 
Unlike quantum mechanical systems with a separable Hilbert space, quantum fields do not have the Gibbs thermal state $e^{-\beta \hat H}/Z$, where $\beta\coloneqq T^{-1}$ is the inverse temperature and $Z=\Tr[e^{-\beta \hat H}]$ is the partition function. 
Instead, we use the so-called Kubo-Martin-Schwinger (KMS) state \cite{Kubo1957thermality, Martin-Schwinger1959thermality}, which can be considered a generalization of the Gibbs state. 
Let $\rho_\beta$ be the KMS state of the scalar field. 
The corresponding Wightman function, $W(\sx, \sx')=\Tr[ \rho_\beta \hat \phi(\sx) \hat \phi(\sx') ]$ can be written as \cite{simidzija2018harvesting}
\begin{align}
    W\ts{th}(\sx, \sx')
    &=
        W\ts{M}(\sx, \sx')
        + 
        W_\beta(\sx, \sx')\,,\label{eq:thermal Wightman}
\end{align}
where $W\ts{M}(\sx, \sx')$ is the Wightman function in the Minkowski vacuum given by \eqref{eq:Wightman in Minkowski} and $W_\beta(\sx, \sx')$ is the contribution coming from the thermality, which reads (for a massless scalar field in $(3+1)$-dimensions)
\begin{align}
    W_\beta (\sx, \sx')
    &=
        \int \dfrac{ \dd^3 k }{ (2\pi)^3 2 \kk  }
        \dfrac{ e^{ -\ii \kk (t-t') + \ii \bk \cdot(\bx - \bx') } + \text{c.c.} }{ e^{ \beta \kk } -1 }\,.
\end{align}

From this Wightman function, one can calculate the elements in the density matrix. 
Let $\mathcal{M}^{\text{th}}$ and $\mathcal{L}^{\text{th}}_{ij}$ denote corresponding elements in the density matrix when the detectors are at rest in a thermal quantum field. 
Employing the concurrence 
\begin{equation}
\mathcal{C}\ts{AB}^{\text{th}}\coloneqq 2 \max \{ 0,\, |\mathcal{M}^{\text{th}}|-\sqrt{ \mathcal{L}\ts{AA}^{\text{th}} \mathcal{L}\ts{BB}^{\text{th}} } \}
\end{equation}
as a measure of entanglement,   for temperatures satisfying $T_1 < T_2$, one can analytically show \cite{simidzija2018harvesting} that $\mathcal{C}^{\text{th}}\ts{AB}(T_1)> \mathcal{C}^{\text{th}}\ts{AB}(T_2)$, namely, the amount of entanglement between two inertial detectors in a thermal bath monotonically decreases with temperature. 
This is distinct from the case of   uniformly accelerating detectors   \cite{Liu:2021dnl}, where entanglement is either   enhanced before vanishing at   high temperature or monotonically decreases, depending on  parameters such as the energy gap $\Omega$. 
It is difficult to analytically show the behavior of quantum mutual information due to its logarithmic definition. However one can numerically check that the mutual information in a thermal bath monotonically increases with temperature \cite{simidzija2018harvesting}. whereas our result (Fig.~\ref{fig:MI different accelerations}) shows  similar  behavior to that in the entanglement harvesting scenario with accelerating detectors \cite{Liu:2021dnl}.

Although an accelerating single detector experiences a thermal bath, two detectors exhibit remarkably different behavior. 
This can be explained by looking at their Wightman functions \eqref{eq:thermal Wightman} and \eqref{eq:para two detector Wightman}. 
One can examine this difference by, for example, performing a series expansion around $T=0$: 
\begin{align}
    &W\ts{th}(\sx, \sx')
    =
        W\ts{M}(\sx, \sx')
        + 
        \dfrac{T^2}{12}
        + \mathcal{O}(T^4)\,, \label{eq:thermal Wightman expansion}\\
    &W_a(\sx, \sx') \notag \\
    &=
        W\ts{M}(\sx, \sx')
        + c_1(\sx, \sx') T\ts{U}
        + c_2(\sx, \sx') T\ts{U}^2
        + \mathcal{O}(T\ts{U}^3)\,,
\end{align}
where the $c_j(\sx, \sx')$ are expansion coefficients that depend on the spacetime points. 
Note that $W\ts{th}(\sx, \sx')$ has an expansion in even-powers of $T$ for arbitrary two points $\sx$ and $\sx'$, and thereby along a single detector trajectory. 
On the other hand, for a single accelerating detector, these functions become $c_1(\sx, \sx')=0$ and $c_2(\sx, \sx')=1/12$, and so $W_a(\sx, \sx')$ reduces to 
\begin{align}
    &W_a(\sx, \sx')
    =
        W\ts{M}(\sx, \sx')
        + \dfrac{T\ts{U}^2}{12}
        + \mathcal{O}(T\ts{U}^4)\,,
\end{align}
which is equivalent to $W\ts{th}(\sx, \sx')$. 
Apparently, the Wightman functions along two trajectories differ, whereas the ones on a single trajectory match. 
It is not so surprising that two distinct Wightman functions give different correlations. 
If we specify the state of the field and the trajectories of the detectors in such a way that two Wightman functions match, the corresponding quantities such as concurrence or transition probability behave in the same way, which is the case for a single accelerating detector.

As an application of this observation, consider two UDW detectors in an expanding universe considered in \cite{Steeg2009, Nambu.correlations.inflation}. 
The line element of the de Sitter spacetime in the planar coordinates is
\begin{align}
    \dd s^2
    &=
        -\dd t^2 + e^{ 2 \kappa t } (\dd x^2 + \dd y^2 + \dd z^2)\,,
\end{align}
where $\kappa$ is the expansion rate of the universe. 
We employ a conformally coupled, massless scalar field in the conformal vacuum. 
In this case, a single inertial detector also sees a thermal bath at temperature $T\ts{GH}\coloneqq \kappa /2\pi$ (the Gibbons-Hawking effect \cite{Gibbons.Hawking}). 
This can be seen from the Wightman function \cite{birrell1984quantum},
\begin{align}
    &W\ts{dS}(\sx, \sx')
    =
        -\dfrac{1}{4\pi^2}
        \dfrac{1}{ \dfrac{ \sinh^2(\pi T\ts{GH} \Delta t-\ii\epsilon) }{ \pi^2 T\ts{GH}^2 } - e^{ 2\pi T\ts{GH} \Delta_+ t } L^2 }\,, \\
    &(\Delta t\equiv t-t' ,\, \Delta_+ t\equiv t+t')\,, \notag 
\end{align}
by pulling it back to a single inertial trajectory, $L=0$,  which yields the same form as \eqref{eq:para single Wightman}.

Nevertheless, a series expansion of $W\ts{dS}(\sx, \sx')$ around $T\ts{GH}=0$ reads
\begin{align}
    &W\ts{dS}(\sx, \sx') \notag \\
    &=
        W\ts{M}(\sx, \sx')
        + c_1^{\text{dS}} (\sx, \sx') T\ts{GH}
        + c_2^{\text{dS}} (\sx, \sx') T\ts{GH}^2
        + \mathcal{O}(T\ts{GH}^3)\,,
\end{align}
and this obviously differs from $W\ts{th}(\sx, \sx')$ in \eqref{eq:thermal Wightman expansion}.

In summary, harvested correlations do not necessarily show the same behavior even if two scenarios give the same transition probability. 
This is simply because the Wightman functions are different in general between distinct spacetime points on different trajectories.
These quantities, including transition probability, show identical features if the Wightman functions in two scenarios happen to be the same.

\section{Conclusion}
\label{sec: conclusion}

We have investigated the harvesting protocol for mutual information with two uniformly accelerating detectors in the Minkowski vacuum. 
Our main purpose was to fill in the missing piece of the correlation harvesting protocol for situations in which a single detector perceives a thermal bath. These include a uniformly accelerating detector, a static detector in a black hole spacetime, and an inertial detector in de Sitter spacetime.

As with the entanglement harvesting scenario revisited recently \cite{Liu:2021dnl}, we considered three types of acceleration scenarios:  parallel, anti-parallel, and perpendicular. 
We found that, as for the entanglement harvesting case, acceleration can enhance mutual information for 
certain 
detector separations and energy gaps. 
Moreover, it asymptotically vanishes as the acceleration (equivalently, the Unruh temperature) increases. 
This is in contrast to the case of two inertial detectors in a thermal bath \cite{Brown2013harmonic, simidzija2018harvesting}, where mutual information monotonically increases with the bath temperature. 
We have also looked into the energy gap dependence and found that there is a range of energy gaps in which either classical correlations or nondistillable entanglement can be extracted from the field.

Our analysis provides a complete picture of correlation harvesting with thermalized detectors. 
The take-home lesson is that the temperature dependence of harvested correlations differs among different scenarios even when a single detector responds in the same manner for each. 
This is not surprising since the properties of harvested correlations depend on the Wightman function (a two-point correlation function of a quantum field) $W(\sx, \sy)= \Tr [ \rho_\phi \hat \phi(\sx) \hat \phi(\sy) ]$, where $\rho_\phi$ is the state of the field $\hat \phi$, which is in general different among different systems. 
Given the state of the field $\rho_\phi$ and the trajectories of detectors, if the functional forms of the Wightman function are different, then one should expect different results for correlation harvesting. 
Nevertheless, if the functional forms of $W(\sx, \sy)$ happen to be identical, then the harvested correlations would show the same behavior. 
In the case of inertial detectors in a thermal bath in \cite{simidzija2018harvesting} and our uniformly accelerating detectors in the Minkowski vacuum, their Wightman functions are different, and so the temperature dependence of entanglement as well as mutual information differs. 
However, if we look at a single detector, the functional forms of the Wightman functions along this trajectory are identical, which is the reason that a single detector undergoing uniform acceleration sees thermality as if it is immersed in a thermal bath at temperature $T\ts{U}=a/2\pi$.

\section*{Acknowledgments}
K. G.-Y. would like to thank Erickson Tjoa for his helpful advice. 
This work was supported in part by the Natural Sciences and Engineering Research Council of Canada.

% \begin{widetext}

% \appendix
% \section{Derivation of $\mathcal{L}\ts{AB}$}\label{app:Derivation of Lij}

% \end{widetext}

\bibliography{ref}

\end{document}